\documentclass[english,12pt,displaymath,a4paper]{article}
\usepackage{amsfonts}
\usepackage{amsmath}
\usepackage{stmaryrd}
\usepackage[T1]{fontenc}
\usepackage{babel}
\usepackage{theorem}
\usepackage{enumerate}
\usepackage{eufrak}
\usepackage{euscript}
\usepackage{newlfont}
\usepackage{dsfont}
\usepackage{amssymb}
\usepackage{tabularx}
\usepackage{eurosym}

\usepackage{epsfig}
\usepackage{graphicx}
\usepackage{indentfirst}
\usepackage{graphics}
\newtheorem{notations}{Notations}

\newcommand{\E}{E}

\begin{document}
\title{The nature of price returns during periods of high market activity}
\author{K. Al Dayri\footnote{CMAP, Ecole Polytechnique, CNRS-UMR 7641, 91128 Palaiseau, France.}, E. Bacry$^{*}$, J.F. Muzy\footnote{CNRS-UMR 6134, Universit\'e de Corse, 20250 Corte, France}$~^{,*}$.}

\maketitle
\begin{abstract}
By studying all the trades and best bids/asks of ultra high frequency
snapshots recorded from the order books of a basket of 10 futures assets, we
bring qualitative empirical evidence that the impact of a single trade
depends on the intertrade time lags. We find that when the trading rate becomes
faster, the return variance per trade or the impact, as measured by the price
variation in the direction of the trade, strongly increases.
We provide evidence that these properties persist at coarser
time scales. We also show that the spread value is an increasing
function of the activity. This suggests
that order books are more likely empty when the trading rate is high.
\end{abstract}

\vskip 7cm
{\small \noindent This research is part of the Chair {\it Financial Risks} of the {\it Risk
Foundation}.

\vskip .2cm

\noindent The financial data used in this paper have been  provided by the company {\em QuantHouse
EUROPE/ASIA}, http://www.quanthouse.com.}

\newpage

\section{Introduction}
\label{sec:1}

During the past decade, the explosion of the amount
of available data associated with electronic markets has permitted
important progress in the description of price fluctuations at the microstructure level.
In particular the pioneering works of Farmer's group  \cite{FarmerLillo2004LongMemory,FarmerLillo2004CauseLargePriceChanges,FarmerLillo2005MoreToVolatility,FarmerLillo2006PriceImpact} and Bouchaud et al.
\cite{Bouchaud2004Molasses,BouchaudGefen2004Fluctuations,Bouchaud2009ImpctOrderBookEv}
relying on the analysis of order book data,
has provided new insights in the understanding of the complex mechanism
of price formation (see e.g \cite{BouchaudFarmerLillo2008} for a recent review).
A central quantity in these works and in most approaches
that aim at modeling prices at their microscopic level, is the market
impact function that quantifies the average response of prices to ``trades''.
Indeed, the price value of some asset is obtained from its cumulated variations
caused by the (random) action of sell/buy market orders.
In that respect, the price dynamics is formulated as a discrete "trading
time" model like:
\begin{equation}
\label{Eq:TransactionTime}
    p_n=\sum_{i<n}G(n-i,V_i)\varepsilon_{i} + diffusion
\end{equation}
where $n$ and $i$ are transaction ``times", i.e., integer indices of market orders.
$V_i$ is the quantity traded at index $i$, $\varepsilon_i$ is the sign of the $i^{th}$ market order
($\varepsilon_i=-1$ if selling and $\varepsilon_i=+1$ if buying).
The function $G(k,V)$ is the bare impact corresponding to the average impact after $k$ trades
of a single trade of volume $V$.
Among all significant results obtained within such a description, one
can cite the weak dependence of impact on the volume of market orders, i.e.,
$G(n,V) \sim G(n) \ln V$, the long-range correlated nature of the sign of the consecutive trades $\varepsilon_i$ and
the resulting non-permanent
power-law decay of impact function $G(n)$ \cite{BouchaudFarmerLillo2008}.
Beyond their ability to reproduce most high frequency stylized facts,
models like \eqref{Eq:TransactionTime} or their continuous counterparts \cite{Almgren2005directestimation}
have proven to be extremely interesting
because of their ability
to control the market impact of a given high frequency
strategy and to optimize its execution cost \cite{Gatheral2009}.

Another well known stylized fact that characterizes price
fluctuations is the high intermittent nature of volatility. This feature
manifests at all time scales, from intradaily scales where periods of
intense variations are observed, for instance, around publications of important
news to monthly scales \cite{BBoPo}. Since early works of Mandelbrot and Taylor \cite{Mantay67},
the concept of subordination by a trading or transaction clock that maps the physical time
to the number of trades (or the cumulated volume) has been widely used in empirical
finance as a way to account for the volatility intermittency.
The volatility fluctuations simply reflects the huge variations of the activity.
The observed intradaily seasonal patterns \cite{DacGenMulOlPic01} can be explained
along the same line. Let us remark that according to the model \eqref{Eq:TransactionTime},
the physical time does not play any role in the way the
market prices vary from trade to trade. This implies notably that
the variance per trade (or per unit of volume traded) is constant and therefore
that the volatility over a fixed physical time scale, is only dependent on the number of
trades.

The goal of this paper is to critically examine this underlying assumption associated with the
previously quoted approaches, namely the fact that the impact of a trade does not depend in any way
on the physical time elapsed since previous transaction.
Even if one knows that volatility is, to a good approximation, proportional
to the number of trades within a given time period (see Section \ref{sec:4}), we aim
at checking to what extent this is true. For that purpose we use a database which includes all the trades
and {\em level 1} (i.e., best ask and best bid) ultra high-frequency snapshots recorded from the order books of
a basket of 10 futures assets. We study the statistics of return variations associated
to one trade conditioned by the last intertrade time.
We find that the variance per trade (and the impact per trade) increases as the speed of trading
increases and we provide plausible interpretations to that. We check that these features
are also observed on the conditional spread and impact. Knowing that the spread is a proxy to the fullness
of the book and the available liquidity \cite{BouchaudWyart2007SpreadImpactVol}, we suspect that in high activity periods
the order books tend to deplete. These "liquidity crisis" states
would be at the origin of considerable amounts of variance not accounted
for by transaction time models.

The paper is structured as follows: In Section \ref{sec:3} we describe the futures data we used
and introduce some useful notations. In Section \ref{sec:4}, we study the variance of price increments
and show that if it closely follows the trading activity, the variance per trade over some fixed time
interval is not constant and increases for strong activity periods. Single
trade variance of midpoint prices conditioned to the last intertrade duration
are studied in Section \ref{sec:5}.
We confirm previous observations made over a fixed time interval
and show that, as market orders come
faster, their impact is greater.
We also show that, for large tick size assets, the variations of volatility for small intertrade times translates essentially
on an increase of the probability for a trade to absorb only the first level of the book (best bid or best ask).
There is hardly no perforation of the book on the deeper levels.
In Section \ref{sec:6} we show that the single trade observations can be reproduced at coarser scales
by studying the conditional variance and impact over 100 trades. We end the section
by looking at the spread conditioned to the intertrade durations. This
allows us to confirm that in period of high activity, the order book tends to empty itself
and therefore the increase in the trading rate corresponds to  a local liquidity crisis.
Conclusions and prospects are provided in Section \ref{sec:7}.

\section{Data description}
\label{sec:3}
In this paper, we study highly liquid futures data, over two years during the period ranging from 2008/08 till 2010/03. We use data of ten futures on different asset classes that trade on different exchanges. On the EUREX exchange (localized in Germany) we use the futures on the DAX index (DAX) and on the EURO STOXX 50 index (ESX), and three interest rates futures: 10-years Euro-Bund (Bund), 5-years Euro-Bobl (Bobl) and the 2-years Euro-Schatz (Schatz).  On the CBOT exchange (localized in Chicago), we use the futures on the Dow Jones index (DJ) and the 5-Year U.S. Treasury Note  Futures (BUS5). On the CME (also in Chicago), we use the forex EUR/USD futures (EURO) and the the futures on the SP500 index (SP). Finally we also use the Light Sweet Crude Oil Futures (CL) that trades on the NYMEX (localized in New-York). As for their asset classes, the DAX, ESX, DJ, and SP are equity futures, the Bobl, Schatz, Bund, and BUS5 are fixed income futures, the EURO is a foreign exchange futures and finally the CL is an energy futures.\\
The date range of the DAX, Bund and ESX spans the whole period from 2008/08 till 2010/03, whereas, for all the rest, only the period ranging from 2009/05 till 2010/03 was available to us. For each asset, every day, we only keep the most liquid maturity (i.e., the maturity which has the maximum number of trades) if it has more than 5000 trades, if it has less, we just do not consider that day for that asset. Moreover, for each asset, we restrict the intraday session to the most liquid hours, thus for instance,  most of the time, we close the session at settlement time and open at the outcry hour (or what used to be the outcry when it no longer exists). We refer the reader to Table 1 for the total number of days considered for each asset (column $D$), the corresponding intraday session and the average number of trades per day.
It is interesting to note that we have a dataset with a variable number of trading days (around 350 for the DAX, Bund and ESX, and 120 for the rest) and a variable average number of orders per day, varying from 10 000 trades per day (Schatz) to 95 000 (SP).
Our data consist of {\em level 1} data : every single market order is reported along with any change in the price or in the quantity at  the best bid or the best ask price. All the associated timestamps are the timestamps published by the exchange (reported to the millisecond). \\
It is important to point out that, since when one market order hits several limit orders it results in several trades being reported, we chose to aggregate together all such transactions and consider them as one market order. We use the sum of the volumes as the volume of the aggregated transaction and as for the price we use the last traded price. In our writing we freely use the terms transaction or trade for any transaction (aggregated or not). We are going to use these transactions as our "events", meaning that all relevant values are calculated at the time of, or just before such a transaction. As such, we set the following notations:

\begin{notations} \label{notations:1}
For every asset, let $D$ be the total number of days of the considered period. We define:
\begin{enumerate}[(i)]
    \item $N_k, k\in\{1\ldots D\}$ the total number of trades on the $k^{th}$ day\label{n1}
    \item $t_i$ is the time of the $i^{th}$ trade  ($i\in[1,\sum_k N_k]$) \label{n3}
    \item $b_{t_i}$ and $a_{t_i}$ are respectively the best bid and ask prices right before the $i^{th}$ trade \label{n4} \item $p_{t_i}=\frac{b_{t_i} + a_{t_i}}{2}$ is midpoint price right before the $i^{th}$ trade \label{n5}
    \item $s_{t_i}=a_{t_i} - b_{t_i}$ is spread right before the $i^{th}$ trade \label{n6}
    \item $r_{t_i}=p_{t_{i+1}}-p_{t_{i}}$ is the return caused by the $i^{th}$ trade, measured in ticks \label{n7}
    \item $NT[s,t]=\#\{t_i, s \leq t_i < t\}$ corresponds to the number of trades in the time interval $[s,t]$ \label{n8}
    \item $\E_t[...]$ or $\E_i[...]$ indifferently refers to the historical average of the quantity in between backets, averaging on all the available days and on all the trading times $t=t_i$. The quantity is first summed up separately on each day (avoiding returns overlapping on 2 consecutive days), then the so-obtained results are summed up and finally divided by the total number of terms in the sum.
    \label{n9}
\end{enumerate}
\end{notations}
Let us note that in the whole paper, we will consider that the averaged returns are always 0, thus we do not include any mean component in the computation of the variance of the returns.
\begin{table}[h]
\begin{small}
     \label{tab:Stats1}
    \centering % centering table
    \begin{tabular}{lcrcccccc} % creating 10 columns
    \hline\hline % inserting double-line
    Futures & Exchange & Tick Value & D & Session & \# Trades/Day  & 1/2-$\eta$ & $P_0$ & $P_{=}$\\ [0.5ex]
    \hline % inserts single-line
 DAX    & EUREX     & 12.5\euro   & 349 & 8:00-17:30 & 56065  & 0.082 & 49    &  67.9\\
    CL     & NYMEX     & 10\$        & 127 & 8:00-13:30 & 76173  & 0.188 & 72.8  &  79.8\\
    DJ     & CBOT      & 5\$         & 110 & 8:30-15:15 & 36981  & 0.227 & 72.6  &  92.2\\
    BUS5   & CBOT      & 7.8125\$    & 126 & 7:20-14:00 & 22245  & 0.288 & 81.6  &  95.1\\
    EURO   & CME       & 12.5\$      & 129 & 7:20-14:00 & 42271  & 0.252 & 79.5  &  95.2\\
    Bund   & EUREX     & 10\euro     & 330 & 8:00-17:15 & 30727  & 0.335 & 80.9  &  97.6\\
    Bobl   & EUREX     & 10\euro     & 175 & 8:00-17:15 & 14054  & 0.352 & 86.5  &  99.1\\
    ESX    & EUREX     & 10\euro     & 350 & 8:00-17:30 & 55083  & 0.392 & 88.3  &  99.2\\
    Schatz & EUREX     & 5\euro      & 175 & 8:00-17:15 & 10521  & 0.385 & 89.3  &  99.4\\
    SP     & CME       & 12.5\$      & 112 & 8:30-15:15 & 97727  & 0.464 & 96.6  &  99.8\\    \hline % inserts single-line
    \end{tabular}
    \caption{Data Statistics. The assets are listed from top to bottom following the increasing order of the $P_=$ column (see \eqref{Eq:PctLrgSpd}), i.e., from the smaller (top) to the greater (bottom) "perceived" tick size. {\em D} : number of days that are considered. The {\em Tick Value} is the smallest variation (expressed in the local currency) by which a trading price can move. The {\em Session} column indicates the considered trading hours (local time).
    The {\em \# Trades/Day}  is the average of the daily number of trades (i.e., $\sum_{k=1}^D N_k/D$ using Notations \ref{notations:1}). $P_0$ and $P_=$ are defined in equations \eqref{Eq:NullRet} and \eqref{Eq:PctLrgSpd} and reported here in percent.}
    \end{small}
%    \label{tab:Stats2}
\end{table}

\paragraph{"Perceived" Tick Size and Tick Value}
The \textsl{tick value} is a standard characteristic of any asset and is measured in its currency. It is the smallest increment by which the price can move. In all the following, all the price variations will be normalized by the tick value to get them expressed in ticks (i.e., in integers for price variations and half-integers for midpoint-price variations).
As one can see in Table 1, column {\em Tick Value}, our assets have very different tick values.
It is important to note a counter-intuitive though very well known fact : the tick value {\em is not} a good measure of the {\em perceived size} (by pratitionners) of the tick. A trader considers that an asset has a small tick when he "feels" it to be negligible, consequently, he is not averse at all to price variations of the order of a single tick.
For instance, every trader "considers" that the the  ESX index futures has a much greater tick than the DAX index futures though the tick values are of the same orders ! There have been several attempts to quantify the perceived tick size.
Kockelkoren, Eisler and Bouchaud in \cite{Bouchaud2009ImpctOrderBookEv}, write that "large tick stocks are such that the bid-ask spread is almost always equal to one tick, while small tick stocks have spreads that are typically a few ticks". Following these lines, we calculate the number of times (observed at times $t_i$) the spread  is equal to 1 tick:
\begin{equation}\label{Eq:PctLrgSpd}
    P_{=}=\frac{\#\{i, s_{t_i}=1\}}{N}
\end{equation}
 and show the results in Table 1. We classify our assets according to this criterion and find SP to have the largest tick, with the spread equal to 1 $99.8\%$ of the time, and the DAX to have the smallest tick.\\
In a more quantitative approach, in order to quantify the aversion to price changes, Rosenbaum and Robert in \cite{RosenbaumRobert2010UncertaintyZonesModel} give a proxy for the perceived tick size using last traded non null returns time-series. If $N^a_t$ (resp. $N^c_t $) is the number of times a trading price makes two jumps in a row in the same (resp. different) directions, then the perceived tick size is given by $1/2-\eta$ where $\eta$ is defined by
 \begin{equation}\label{Eq:eta}
    \eta=\frac{N^c_t}{2N^a_t}
 \end{equation}
For each asset, we computed $\eta$ for every single day, and average over all the days in our dataset and put the result in the  {\em $1/2-\eta$} column in Table 1. We find that the rankings of the assets using this criterion almost matches the ranking using \cite{Bouchaud2009ImpctOrderBookEv}'s $P_{=}$ criterion
(two slight exceptions being the ESX/Schatz and BUS5/EURO ranking). One interpretation of the $\eta$ based proxy is that if the tick size is large, market participant are more averse to changes in the midpoint price and market makers are happy to keep the spread collapsed to the minimum and the midpoint would only move when it becomes clear that the current price level is unsustainable. To check that, we calculate the number of times (observed at times $t_i$) the return (as defined in notation \ref{notations:1}) is null:
\begin{equation}\label{Eq:NullRet}
P_{0}=\frac{\#\{i, r_{t_i}=0\}}{N}
\end{equation}
 and show the result in Table 1. Again, it approximately leads to the same ranking which has nothing to do with the ranking using Tick Values.

\section{Realized variance versus number of trades}
\label{sec:4}
\begin{figure}[h]
    \centering
        \includegraphics[width=1\linewidth]{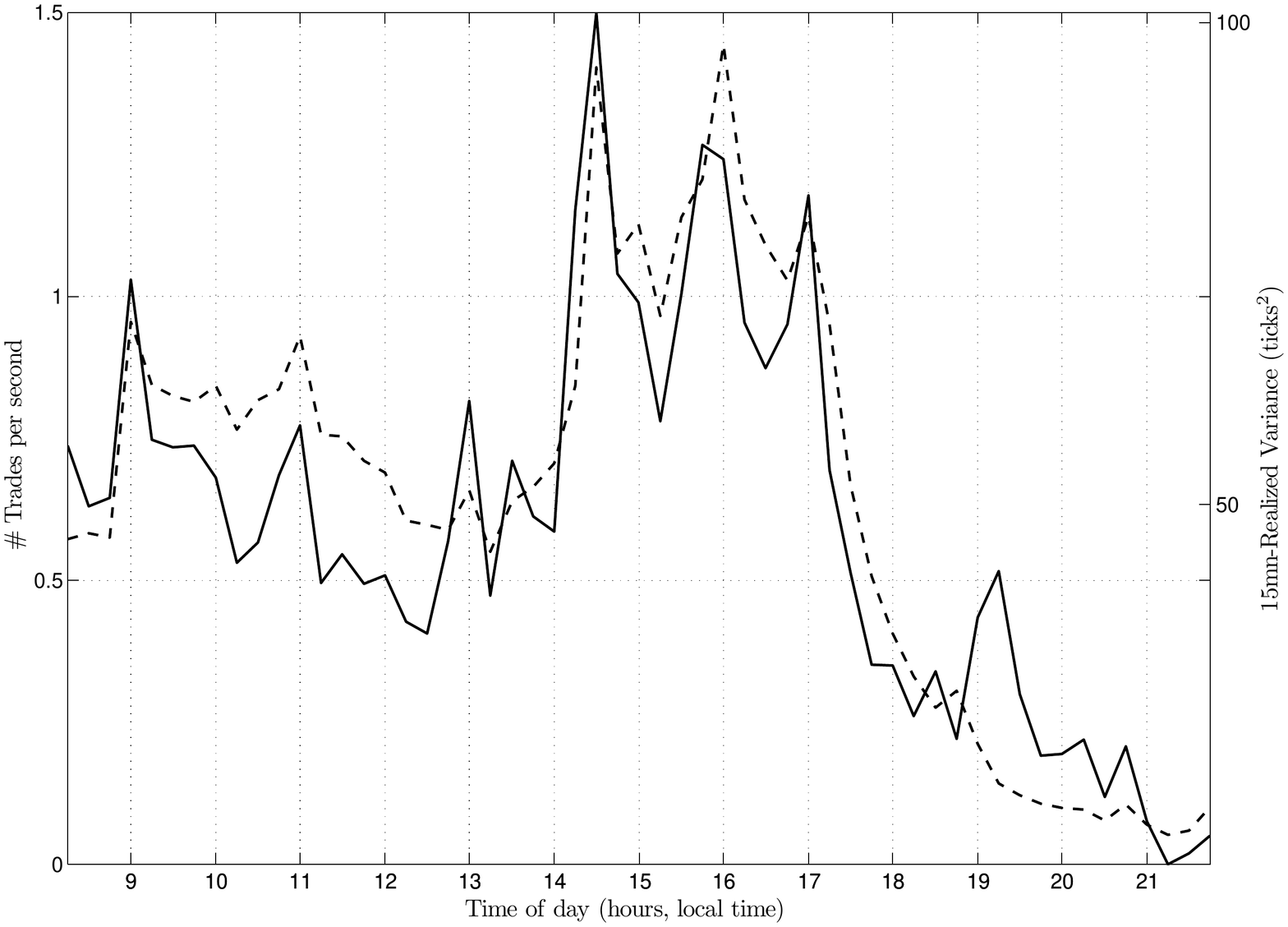}
    \caption{Bund intraday seasonality of both trading rate and volatility (abscissa are expressed in hours, local time). Averages are taken on all available days. Dashed line :  average intraday rate of trading (average number of trades per second) using 15mn bins. Solid line : average 15-minutes-realized variance (estimated summing on 15 1mn-squared returns).}\label{fig:TrdVarAvgPrf-Bund}
\end{figure}

It is widely known that, in a good approximation,  the variance over some period
of time is proportional to the number of trades during that time period
(see e.g. \cite{DacGenMulOlPic01}). Fig. \ref{fig:TrdVarAvgPrf-Bund} illustrates this property on Bund data.
On 15 minutes intraday intervals,
averaging on every single day available,
we look at (dashed curve) the average intraday rate
of trading (i.e., the average number of trades per second) and (solid curve) the average (15 minutes)
realized variance (estimated summing on 1mn-squared returns $(p_{t+1mn}-p_t)^2$).
We see that the so-called intraday seasonality of the variance
is highly correlated with the intraday seasonality of the trading rate \cite{DacGenMulOlPic01}.
\begin{figure}[h]
    \centering
        \includegraphics[width=1\textwidth]{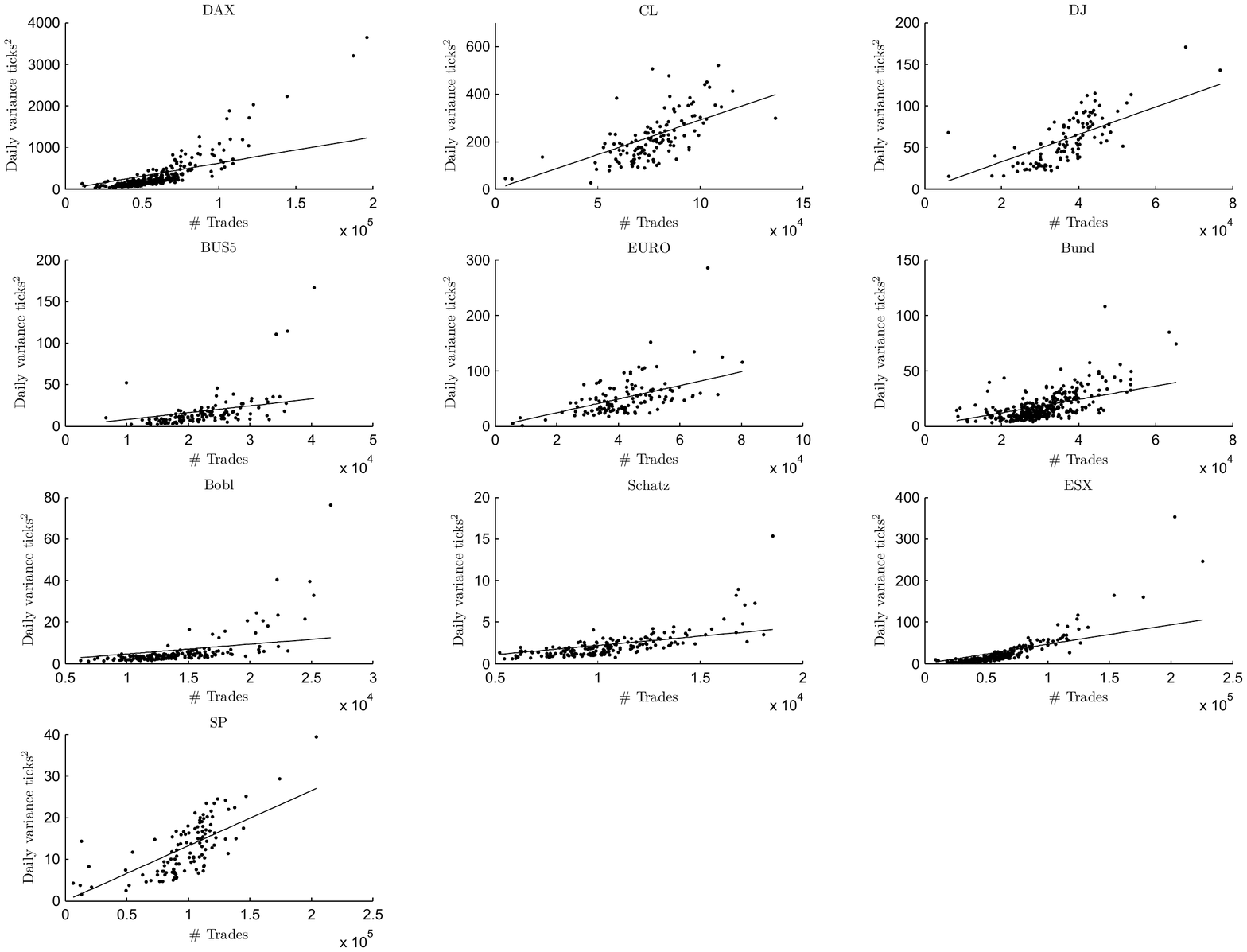}
    \caption{For each asset (in increasing perceived tick size $P_=$) : Daily variance (estimated summing over 5-mn quadratic returns) against daily number of trades. Each dot represents a single day. The solid line is the linear regression line with zero intercept. We see strong linearity between the variance and the number of trades but there seem to be clustering of dots above (resp. below) the solid line for days with high (resp. low) activity.}\label{fig:TradesVsRV}
\end{figure}
%However, the statistics of the trade arrival process are complicated (see \cite{Hautsch2004ModellingIrregularData} for an overview of trade arrival times models, or more %recently \cite{Hewlett2006Clustering, Howard2010SelfExcitedPointProcess}), let us look at the
In order to have more insights, we look at some daily statistics~:
Fig. \ref{fig:TradesVsRV} shows a scatter plot in which each point corresponds
to a given day $k$ whose abscissa is the number of trades within this day, i.e., $N_k$, and the ordinate
is the daily variance (estimated summing over 5-mn quadratic returns) of the same day $k$.
It shows that, despite some dispersion,
the points are mainly distributed around a mean linear trend
confirming again the idea shown in Fig. \ref{fig:TrdVarAvgPrf-Bund} that,
to a good approximation, the variance is proportional to the number of trades. In that respect,
trading time models (Eq. \eqref{Eq:TransactionTime}) should capture most
of the return variance fluctuations through
the dynamics of the transaction rate. However,  in Fig. \ref{fig:TradesVsRV},
the points with high abscissa values (i.e., days with a lot of activity) tend to be located above the linear line,
whereas the ones with low abscissa (low activity) cluster below the linear line,
suggesting that the variance per trade is dependant on the daily intensity of trading.

Before moving on, we need to define a few quantities.
Let $\Delta t$ be an intraday  time scale  and let $N$ be a number of trades. We define
$V(\Delta t,N)$ as the estimated price variance over the scale $\Delta t$ conditioned by the fact that $N$ trades occurred.
Using notations, \ref{notations:1} ({\em\ref{n8}}) and ({\em \ref{n9}}),  from a computational point of view, when $\Delta t=\Delta t_0$ is  fixed and $N$ is varying,
$V(\Delta t=\Delta t_0, N)$ is estimated as:
\begin{equation}
\label{VdtN1}
V(\Delta t=\Delta t_0, N) = \E_{t} \left[(p_{t+\Delta t_0}-p_{t})^2~ |~NT[t,t+\Delta t_0] \in [N-\delta_N,N+\delta_N] \right].
\end{equation}
where $\delta_N$ is some bin size. And, along the same line, when we study $V(\Delta t,N)$ for a fixed $N=N_0$ value
%(e.g. $N=1$ as in Section \ref{sec:5})
over a range of different values of $\Delta t$, one defines a temporal bin size $\delta_{\Delta t}$ and
computes $V(\Delta t,N=N_0)$ as\footnote{Let us point out that we used the index $i-1$ in the condition of \eqref{VdtN2} and not the index $i$ since, for the particular case $N_0= 1$ (extensively used in Section \ref{sec:5}), we want to use a {\em causal} conditionning of the variance. For $N _0$ large enough, using one or the other does not really matter.}
\begin{equation}
\label{VdtN2}
V(\Delta t, N=N_0) = \E_{i} \left[(p_{t_{i+N_0}}-p_{t_i})^2 ~| ~t_{i-1+N_0}-t_{i-1}  \in [\Delta t-\delta_{\Delta t},\Delta t+\delta_{\Delta t}] \right].
\end{equation}
Let us note that, in both cases, the bins are chosen such that each bin involves approximately the same number of terms.
We also define the corresponding conditional variance per trade as:
\begin{equation}
\label{vdtN}
v(\Delta t,N) = \frac {V(\Delta t,N)}{N}.
\end{equation}

\begin{figure}[h]
    \centering
        \includegraphics[width=1\textwidth]{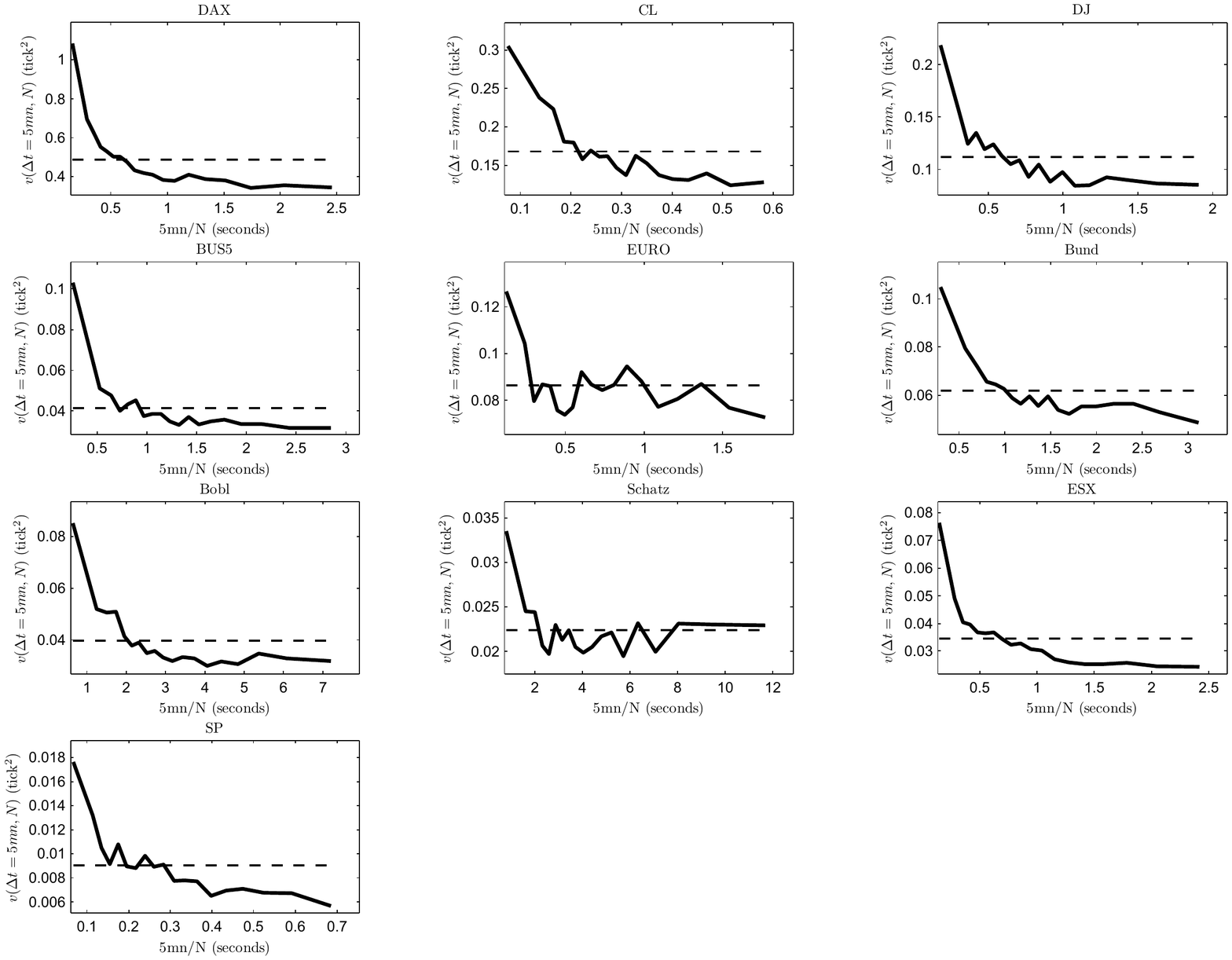}
    \caption{For each asset (in increasing perceived tick size $P_=$), Solid line :
    conditional $v(\Delta t = 5mn,N)$ variance per trade (see \eqref{VdtN1}
    as a function of the average intertrade duration
    $\frac {5mn} N$ when varying $N$. Dashed line : unconditional 5mn-variance per trade.
 The solid line is almost constant for average times above 0.6 seconds,
  and it increases when the trading becomes faster.}\label{fig:DurationVsRvPerTrade5mins}
\end{figure}

In order to test the presence of an eventual non-linear behavior
in the last scatter plots (Fig.~\ref{fig:TradesVsRV}),
we show in Fig. \ref{fig:DurationVsRvPerTrade5mins} the 5-minutes
variance per trade  $v(\Delta t = 5mn,N)$ as a function of the average intertrade duration
$
\frac {5mn} N
$
as $N$ is varying. We clearly  see that
the estimated curve (solid line)
is below the simple average variance (dashed line)
for large intertrade durations and above the average variance when the trades are
less than $600$ milliseconds apart. Note that we observed a similar behavior
for most of the futures suggesting a universal behavior.

To say that the realized variance is proportional to the number of trades is
clearly a very good approximation as long as the trading activity is not too high
as shown both on a daily scale in  Fig. \ref{fig:TradesVsRV} and on a
5mn-scale in Fig. \ref{fig:DurationVsRvPerTrade5mins}.
However, as soon as the trading activity is high
(e.g., average intertrade duration larger than $600ms$  on a 5mn-scale),
the linear relationship seems to be lost. In the next section we will focus on the impact
associated with a single trade.
\begin{figure}[h]
   \centering
       \includegraphics[width=1\textwidth]{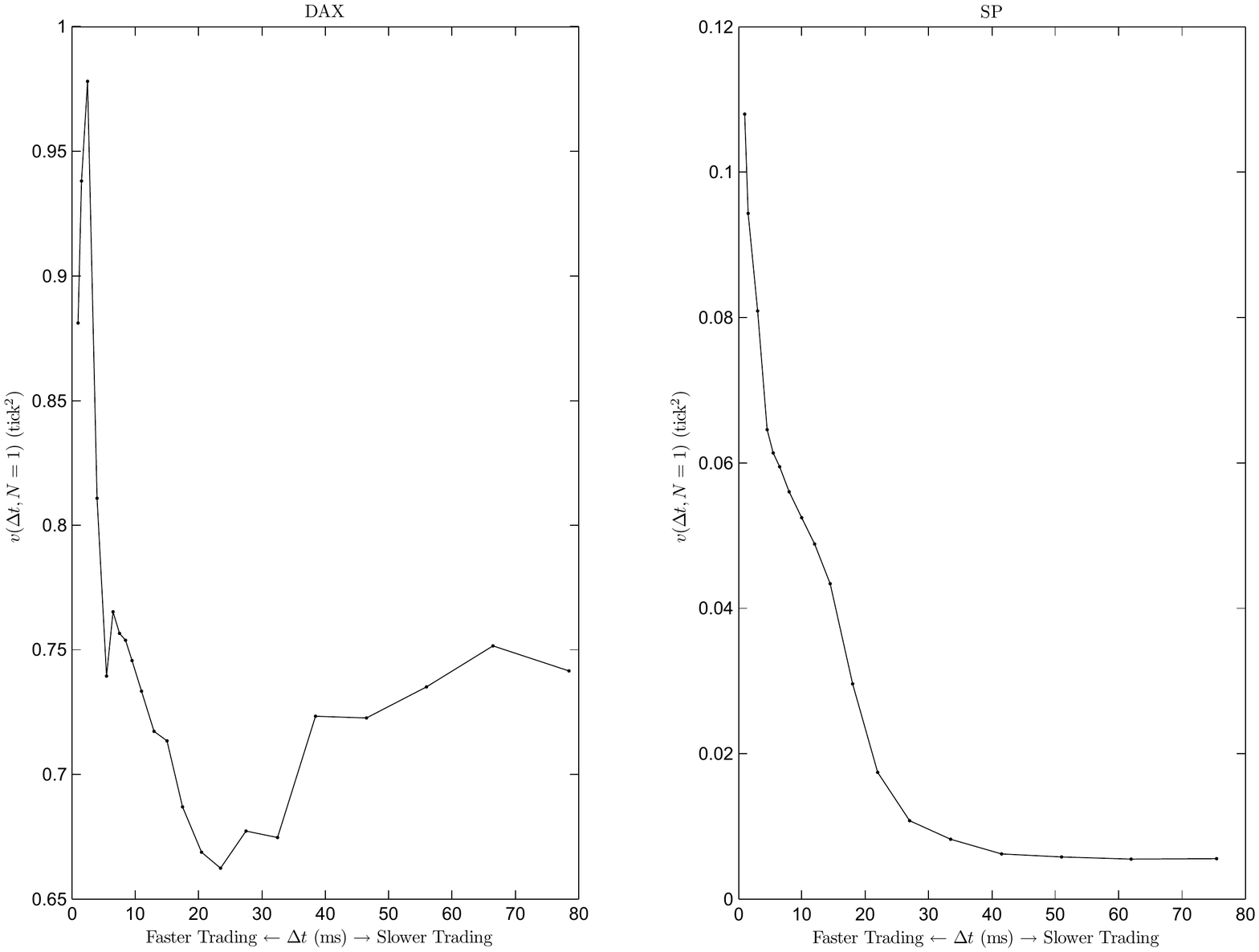}
   \caption{$v(\Delta t,N=1)$ as a function of $\Delta t$
   over very short $\Delta t$'s for DAX and SP.
   The variance per trade increases dramatically below a certain $\Delta t$.}\label{fig:VarCondDt-DAX-SP-N-1-UZ}
\end{figure}

\section{Single trade impact on the midpoint price}
\label{sec:5}

In this section, we will mainly focus on the impact of a trade $i$, and more specifically
on the influence of its arrival time $t_i$ on the return $r_{t_i} = p_{t_{i+1}}-p_{{t_i}}$.
In order to do so, it is natural to consider the return $r_{t_{i}}$ conditioned
by $t_i-t_{i-1}$, the time elapsed since previous transaction. We want to be able to
answer questions such as : how do compare the impacts of the $i$th trade
depending on the fact that it arrived right after or long after the previous trade ?
Of course, in the framework of trading time models this question has a very simple answer : the impacts are the same !
Let us first study the conditional variance of the returns.

\subsection{Impact on the return variance}
In order to test the last assertion, we are
naturally lead to use Eqs \eqref{VdtN2} and \eqref{vdtN} for $N_0 = 1$, i.e,
\begin{equation}
\label{vNdt}
v(\Delta t,N=1) = \E_i\left[r_{{t_i}}^2 ~| ~t_{i}-t_{i-1} \in [\Delta t-\delta_{\Delta t},\Delta t+\delta_{\Delta t}] \right].
\end{equation}
Let us illustrate our purpose on the DAX and the SP futures. They trade on two different exchanges,
(EUREX and CME) and have very different daily statistics (e.g.,
DAX has the smallest perceived tick and SP the largest as one can see in Table 1).
Fig. \ref{fig:VarCondDt-DAX-SP-N-1-UZ} shows for both assets $v(\Delta t,N=1)$ (expressed in squared $tick$)
as a function of $\Delta t$ (in milliseconds).
We notice that both curves present a peak for very small $\Delta t$ and stabilize
around an asymptotic constant value for larger $\Delta t$.
This value is close to $0.7$ ticks$^2$ for the DAX and to $0.005$ ticks$^2$ for the SP.
The peak reaches $0.95$ ($35\%$ above the asymptote)
for the DAX, and $0.1$ ($2000\%$ above the asymptote).
\begin{figure}[h]
    \centering
        \includegraphics[width=1\textwidth]{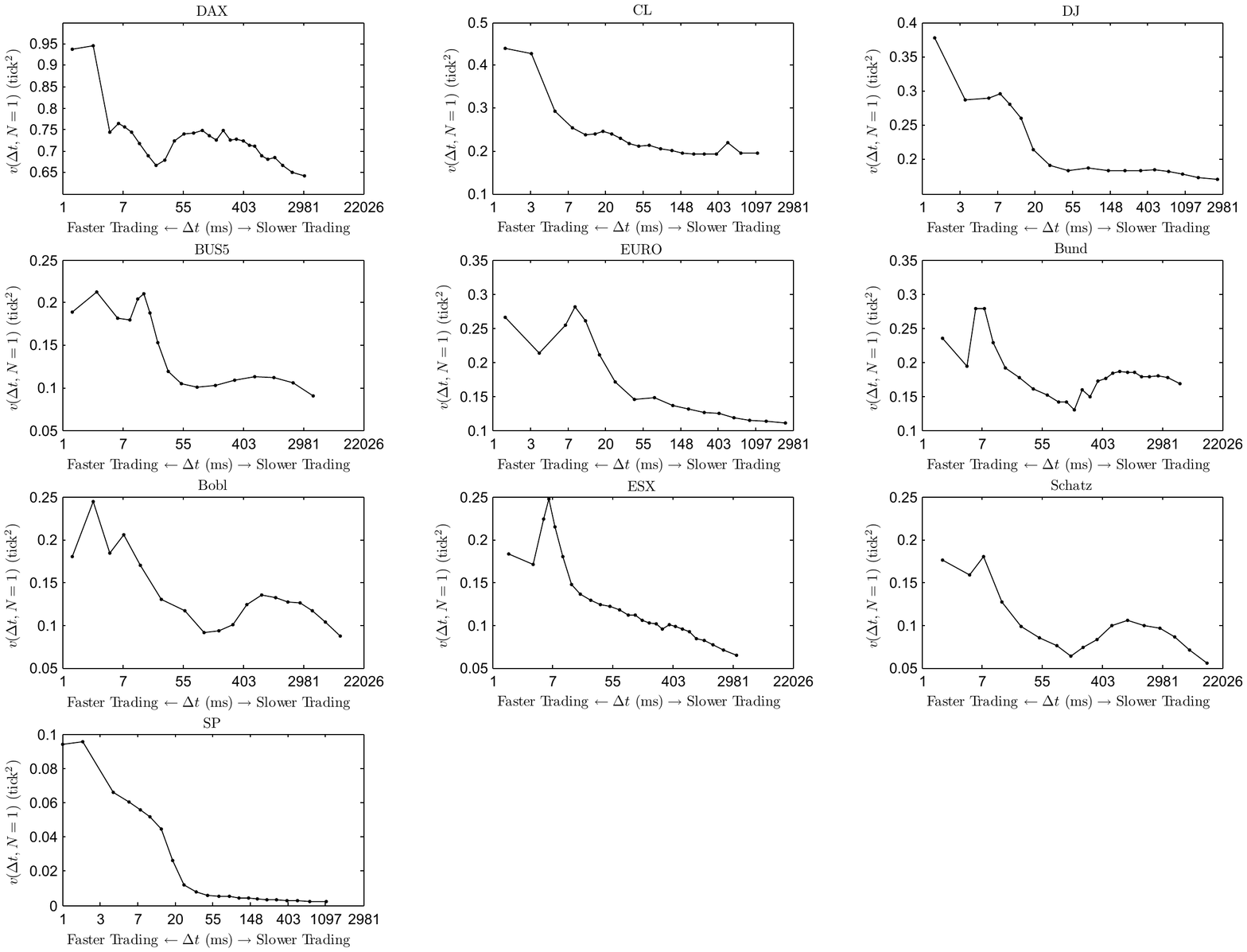}
    \caption{For each asset (in increasing perceived tick size $P_=$), $v(\Delta t,N=1)$
    as a function of $\Delta t$ (logarithm scale). We see an "explosion" of the variance when the trading is getting faster. }\label{fig:VarCondMu-All-N-1}
\end{figure}

Fig. \ref{fig:VarCondMu-All-N-1} switches (for all assets) to a log scale in
order to be able to look at a larger time range.
A quick look at all the assets show that they present a very similar behavior.
One sees in particular for the ESX curve that the variance increases almost
linearly with the rate of trading, and then suffers an explosion as
$\Delta t$ becomes smaller than 20 ms.
The "same" explosion can be qualitatively observed over all assets albeit
detailed behavior and in particular the minimal threshold $\Delta t$ may
vary for different assets.

%\subsection{Impact on the return dynamics???}
%{\bf \Large??? J'aime pas ce titre et je trouve que cette section prend un peu trop de place - je suis embete}
Let us note that the variance $v(\Delta t,N=1)$ as defined by \eqref{vNdt} can be written in the following way:
\begin{equation}
\label{vdecomp}
v(\Delta t,N=1) = P(\Delta t) A(\Delta t),
\end{equation}
where $P(\Delta t)$ is the probability for the return to be non zero conditioned by the intertrade duration  $t_{i}-t_{i-1}=\Delta t$, i.e.,
\begin{equation}
\label{P}
P(\Delta t) = Prob\{r_{{t_i}}\neq  0 ~|~   t_{i}-t_{i-1}\in [\Delta t-\delta_{\Delta t},\Delta t+\delta_{\Delta t}]\}
\end{equation}
and where $A(\Delta t)$ is the expectation of the squared return conditioned
by the fact that it is not zero and by the intertrade duration  $t_{i}-t_{i-1}=\Delta t$, i.e.,
\begin{equation}
\label{A}
A(\Delta t) = \E_i\left[r_{{t_i}}^2 ~|~ r_{{t_i}}\neq  0~{\mbox{and}}~   t_{i}-t_{i-1}\in [\Delta t-\delta_{\Delta t},\Delta t+\delta_{\Delta t}]\right]
\end{equation}
\begin{figure}[h]
    \centering
        \includegraphics[width=1\textwidth]{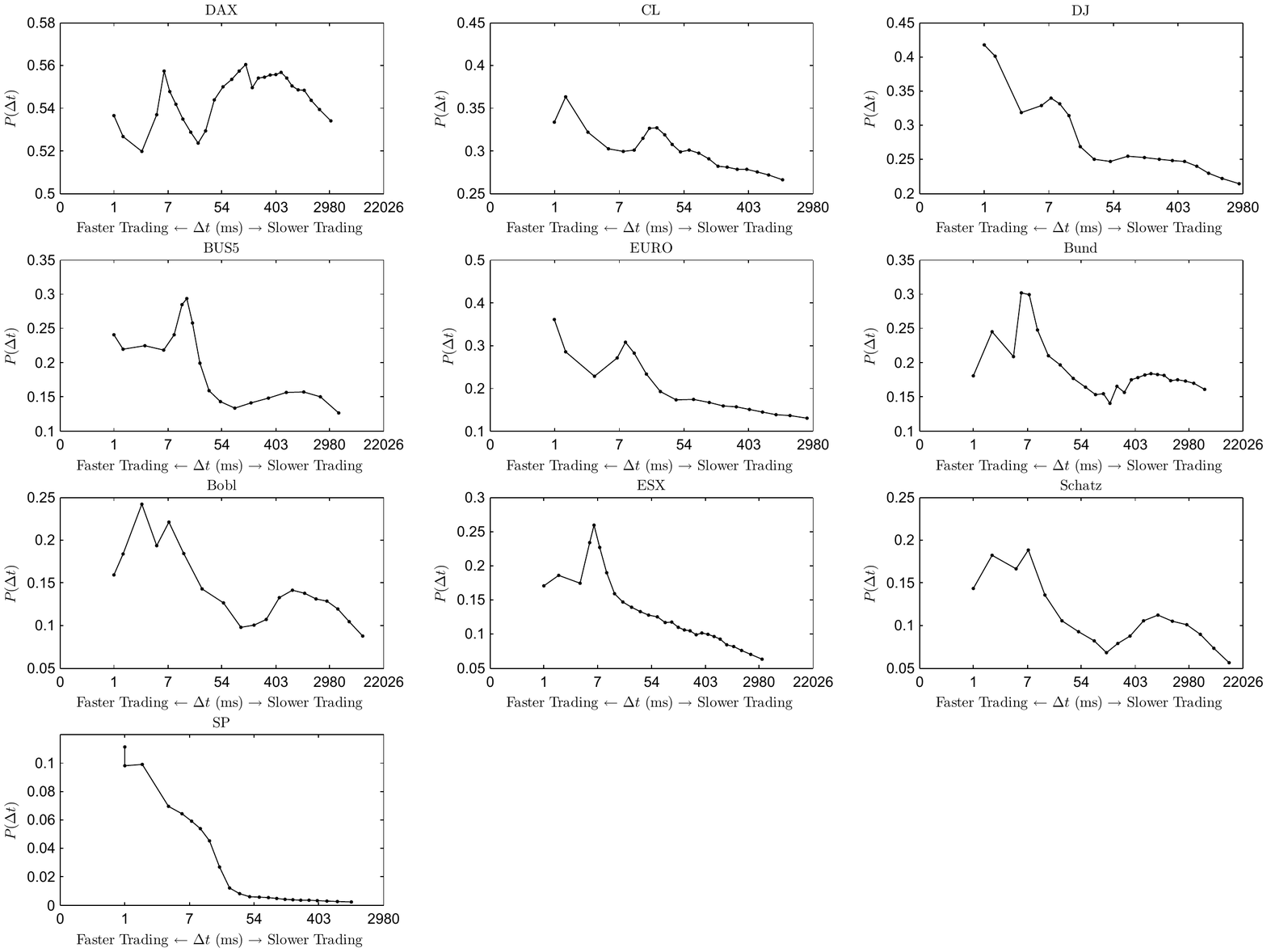}
   \caption{For each asset (in increasing perceived tick size $P_=$),
   Probability $P(\Delta t)$ as defined by \eqref{P} as a function of $\Delta t$.
   We see that the probability of getting a price move increases with market order rate for most assets.}
   \label{fig:ProbaNonNullReturn-All-N-1}
\end{figure}
\begin{figure}[h]
    \centering
        \includegraphics[width=1\textwidth]{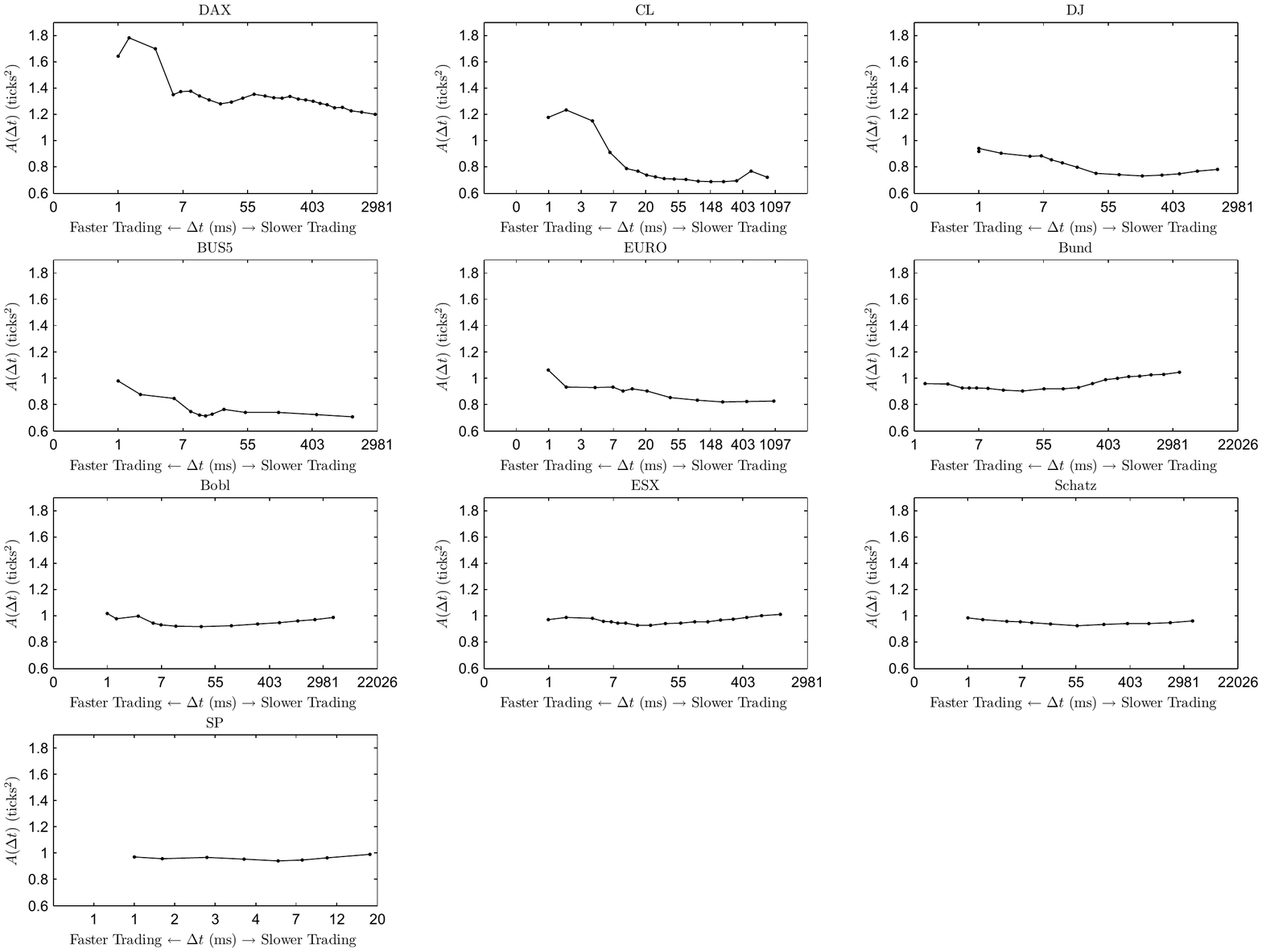}
   \caption{For each asset (in increasing perceived tick size $P_=$),
   Size of Absolute squared returns $A(\Delta t)$ as a function pf $\Delta t$. For very small tick assets, like DAX and CL we see that the absolute size of a return increases with the rate of market orders. This propriety quickly stops being true as the tick increases. The orderbook of a large tick asset is generally much thicker than that of a small tick asset and therefore it is extremely hard to to empty more than one level.}\label{fig:AbsRetMove-All-N-1}
\end{figure}
In short $P(\Delta t)$ is the probability that the midpoint price moves while $A(\Delta t)$ is the squared amplitude of the move when non-zero.
In Fig. \ref{fig:ProbaNonNullReturn-All-N-1}, we have plotted, for all assets, the function $P(\Delta t)$
for different $\Delta t$. One clearly sees that, as the trading rate becomes greater
($\Delta t \rightarrow 0$), the probability to observe a move of the midpoint price increases. One mainly
recovers the behavior we observed for the analog variance plots.
Let us notice that (except for the DAX), the values of the moving probabilities globally
decrease as the perceived ticks $P_=$ increases (for large ticks, e.g. SP, at very low activity this probability
is very close to zero). The corresponding estimated moving squared amplitudes $A(\Delta t)$ are displayed
in Fig. \ref{fig:AbsRetMove-All-N-1}. It appears clearly that, except for the smallest perceived ticks assets (DAX and CL basically), the amplitude can be considered as constant. This can be easily explained~:  large tick assets never make moves larger than one tick while small tick assets are often
``perforated'' by a market order.
One can thus say that, except for very small ticks assets, the variance increase in high trading rate period is mostly caused by the
increase of the probability that a market order absorb only the first level of the book (best bid or best ask). There is hardly no perforation of the book on the deeper levels.

\subsection{Impact on the return}
\begin{figure}[h]
    \centering
        \includegraphics[width=1\textwidth]{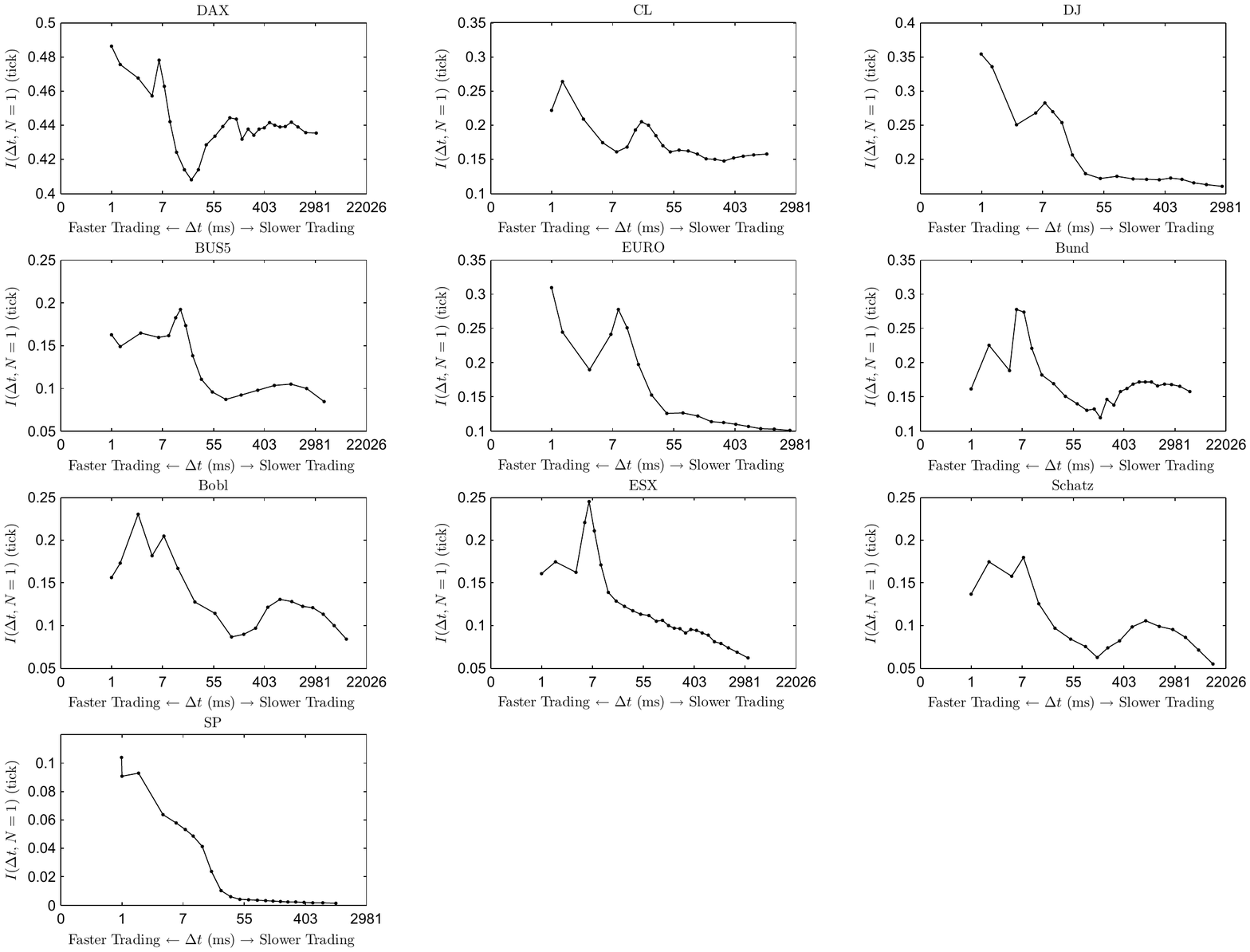}
    \caption{For each asset (in increasing perceived tick size $P_=$), $I(N=1 | \Delta t)$  as defined by \eqref{ImctCondMu} as a function of $\Delta t$. The shape of the curves confirms the idea that the impact is high correlated with the variance per trade.}\label{fig:ImpctCondMu-All-N-1}
\end{figure}

Before moving to the next section, 
let us just look at the direct impact on the return itself, as defined
for instance by \cite{BouchaudFarmerLillo2008}, conditioned by the intertrade time:
\begin{equation}
\label{ImctCondMu}
I(\Delta t,N=1) = \E_i\left[\varepsilon_i r_{{t_i}} | t_{i}-t_{i-1}=\Delta t\right].
\end{equation}
According to \cite{BouchaudWyart2007SpreadImpactVol}, we expect the impact to be correlated with
the variance per trade and therefore for $I(\Delta t)$ to follow a very similar shape
to that of $v(\Delta t, N = 1)$ shown in Fig. \ref{fig:VarCondMu-All-N-1} .
This is confirmed in Fig. \ref{fig:ImpctCondMu-All-N-1} where one sees that, for all assets,
the impact goes from small values for large intertrade intervals to significantly higher values for
small intertrade durations.

\section{From fine to coarse}

\subsection{Large scale conditional variance and impact}
\label{sec:6}
\begin{figure}[h]
    \centering
        \includegraphics[width=1\textwidth]{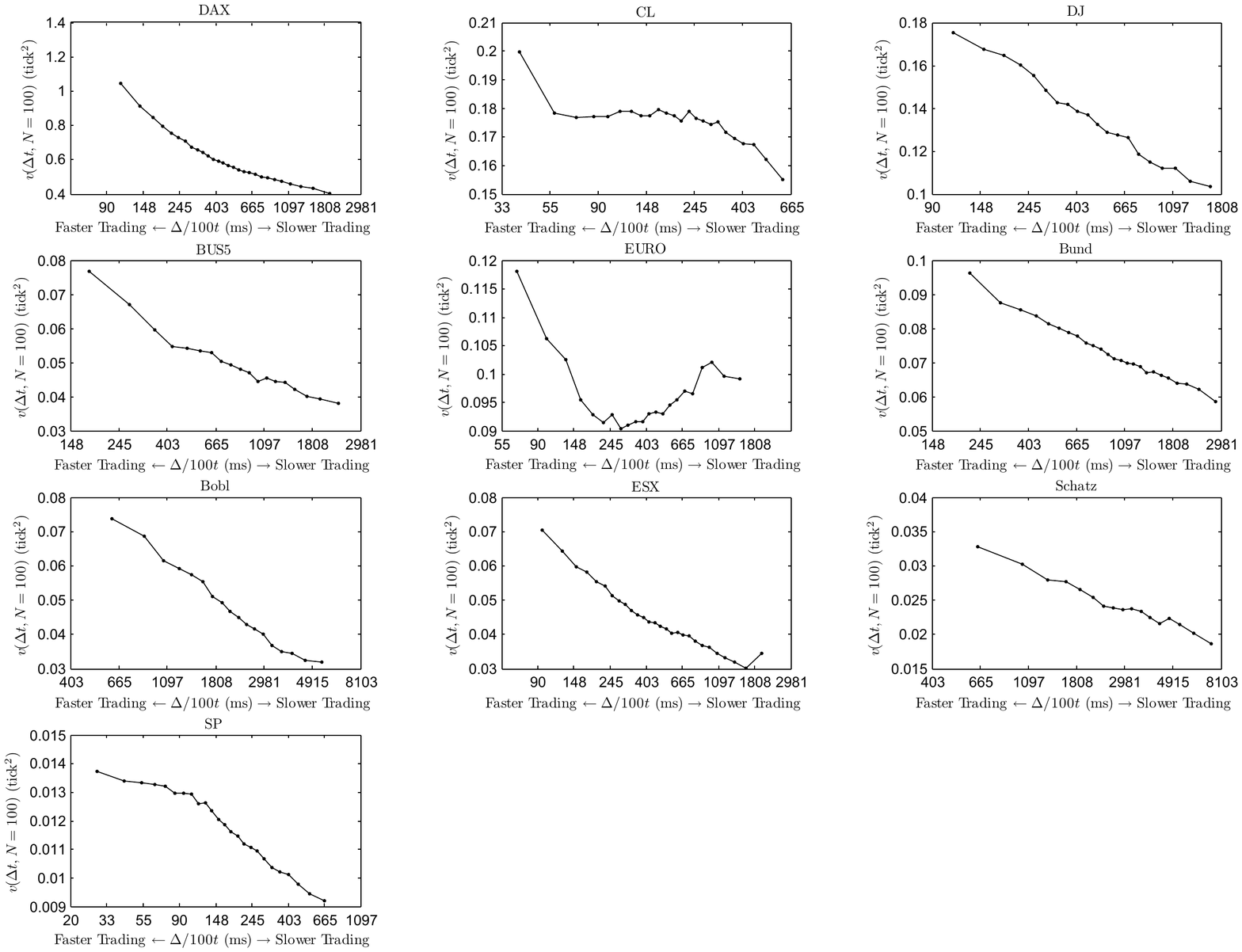}
    \caption{For each asset (in increasing perceived tick size $P_=$),   $v(\Delta t,N = 100)$, as defined by \eqref{vdtN}, as a function of $\Delta t$. Clearly the variance of a speedy $100$ trades is higher than the variance of $100$ slow trades.   }\label{fig:VarNCondNMu-All-N-100}
\end{figure}

%We now move to the see the impact on a coarse scale.We show what happens to the impact of the speed one trade, $200$ trades later by plotting $E(\epsilon_1*(p_{N+1}-%p_1)|t_1-t_0)$ in figure \ref{fig:ImpctCondMu-All-N-200}. The figures show that the impact of a speedy trade propagates and is felt well into the future. Why the impact %propagates into the future and the variance does not, has to do with the autocorrelation of midpoint returns, which we also found to be dependent on the rate of trading. On %the other hand the impact of of a fast trade at time 1 is conserved in $\epsilon_1*(p_{N+1}-p_1)=\epsilon_1*\sum_{i=1}^{N}p_{i+1}-p_i$ and appears at the end.

\begin{figure}[h]
    \centering
        \includegraphics[height=0.7\textwidth]{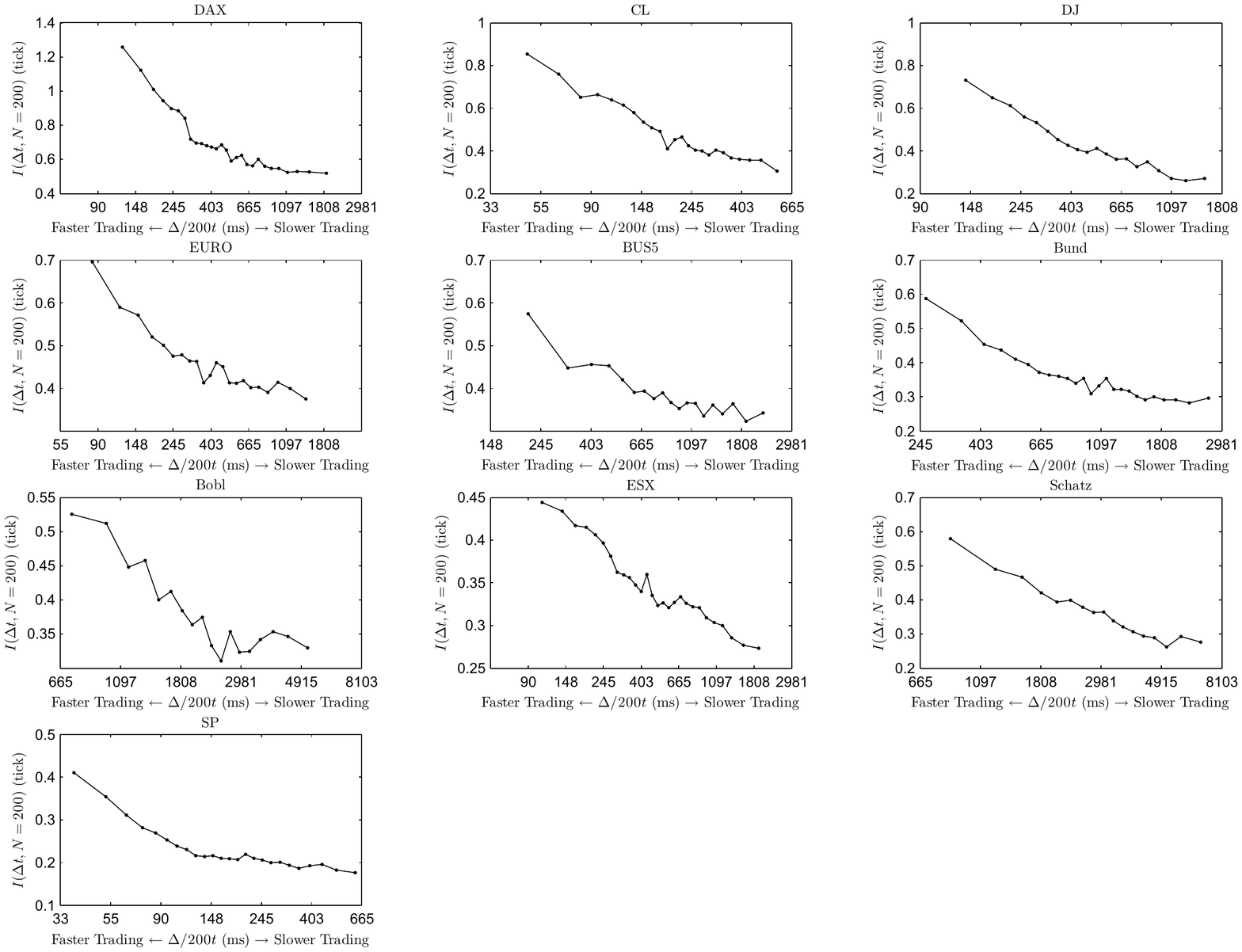}
    \caption{For each asset (in increasing perceived tick size $P_=$),   $I(\Delta t,N = 200)$, as defined by \eqref{ImctCondMuN},  as a function of $\Delta t$.
    The impact of speedy trades propagates well into the future. Even $200$ trades away, one speedy trade has caused more impact than a slower one.}\label{fig:ImpctCondMu-All-N-200}
\end{figure}

One of the key issue associated to our single trade study is the understanding of the consequences
of our findings to large scale return behavior. This question implies the study of (conditional) correlations
between successive trades, which is out of the scope of this paper and will be addressed in a forthcoming work.
However one can check whether the impact or the variance averaged locally over a large number of trades still display
a dependence as respect to the trading rate. Indeed, in Fig. \ref{fig:DurationVsRvPerTrade5mins} we have already seen that
this feature seems to persist when one studies returns over a fixed time (e.g., 5 min) period conditioned by the
mean intertrade duration over this period. Along the same line, one can fix a large $N=N_0$ value and
compute $v(\Delta t,N=N_0)$  and $I(\Delta t,N=N_0)$ as functions of $\Delta t$.
Note that $v(\Delta t,N=N_0)$ is defined in Eq. \eqref{vdtN}
while the aggregated impact can be defined similarly as:
\begin{equation}
\label{ImctCondMuN}
I(\Delta t, N=N_0) = \E_{i} \left[\epsilon_i(p_{t_{i+N_0}}-p_{t_i}) ~| ~t_{i-1+N_0}-t_{i-1}  \in [\Delta t-\delta_{\Delta t},\Delta t+\delta_{\Delta t}] \right].
\end{equation}
In Fig. \ref{fig:VarNCondNMu-All-N-100} and \ref{fig:ImpctCondMu-All-N-200} are plotted respectively the variance $v(\Delta t,N=100)$ and the return impact
$I(\Delta t,N=200)$ as functions of $\Delta t$. One sees that at these coarse scales, the increasing of these two quantities as the activity increases is
clear (except maybe for the variance of the EURO). As compared to single trade curves, the threshold-like behavior are smoothed out and both variance and return impacts go continuously from
small to large values as the trading rate increases.

\subsection{Liquidity decreases when trading rate increases}
One possible interpretation of these results would be that when
the trading rate gets greater and greater, the liquidity tends to decrease, i.e.,
the order book tends to empty.

\begin{figure}[h]
   \begin{center}
       \noindent\makebox[\textwidth]{\includegraphics[width=1\textwidth]{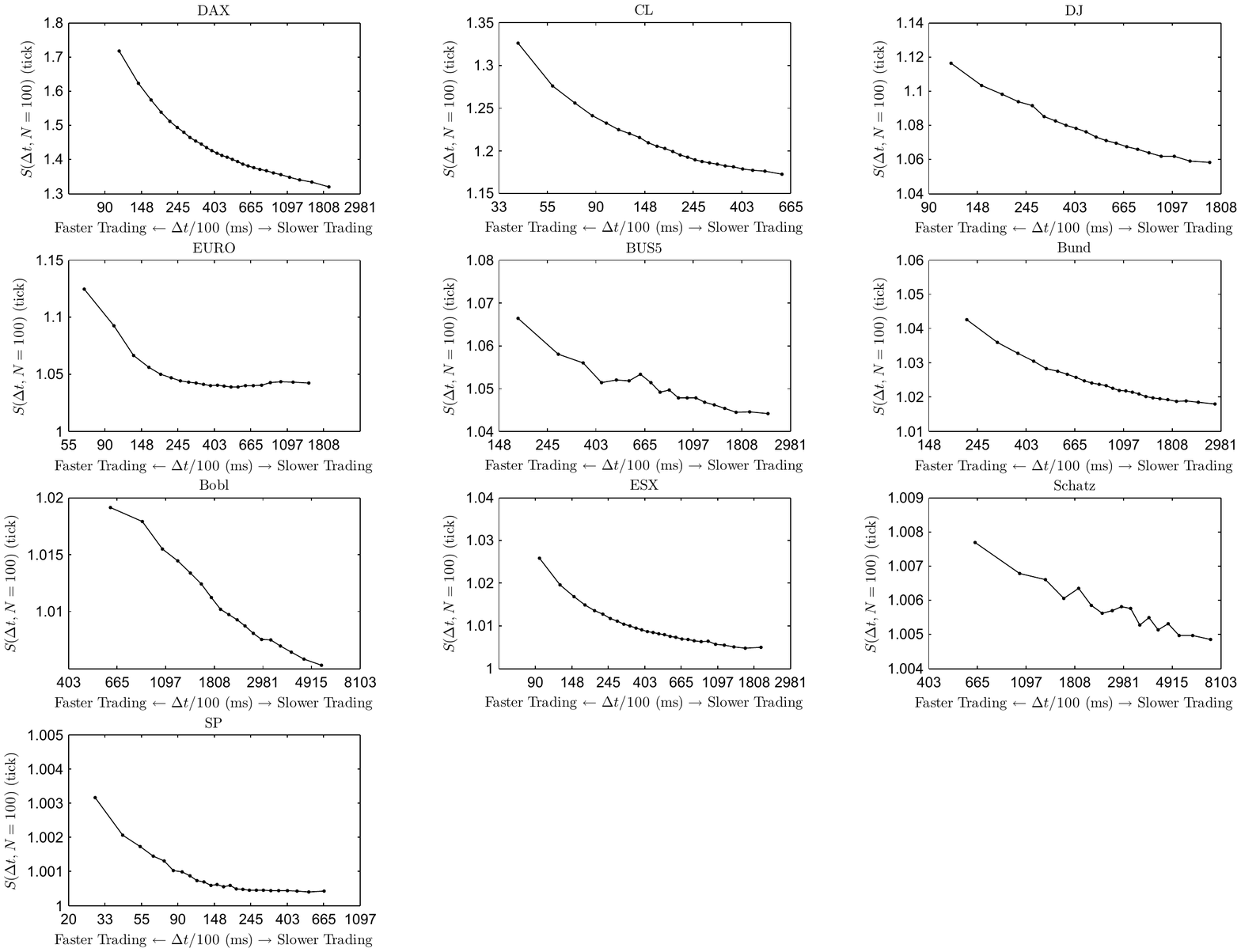}}
   \end{center}
   \caption{For each asset (in increasing perceived tick size $P_=$), $S(\Delta t,N=100)$ as defined in \eqref{spread} as a function of $\Delta t/100$. The form of the curves confirms that there is a strong liquidity decrease when the trading rate is increasing.}\label{fig:SpdCondMu-All-N-1}
\end{figure}

In \cite{BouchaudWyart2007SpreadImpactVol}, the authors mention
that the spread is an indicator of the thinness of the book and
that the distance from the best bid or ask to the next level of the
order book is in fact equivalent to the spread. Moreover, they bring empirical evidence
and theoretical no-arbitrage arguments suggesting that the spread and the variance per trade are strongly correlated.
Accordingly, we define the mean spread over $N$ trades as
\begin{equation}
s_{t_i,N} =  \frac 1  N \sum_{k=0}^{N-1} s_{t_{i+k}},
\end{equation}
and the conditional spread at the fixed scale $N=N_0$ as
\begin{equation}
\label{spread}
S(\Delta t,N=N_0) = \E_i \left[ s_{t_{i},N} ~ | ~t_{i+N}-t_{i} \in [\Delta t-\delta_{\Delta t},\Delta t+\delta_{\Delta t}]\right].
\end{equation}
Fig. \ref{fig:SpdCondMu-All-N-1} displays, for each asset,
$S(\Delta t, N = 100)$ as a function of $\Delta t/100$ (using log scale).
One observes extremely clearly an overall increase of the
spread value with the rate of trading for all assets,
This clearly suggests that the order book is thinner during periods of intense trading.
This seems to be a universal behavior not depending at all on the perceived tick size.

\section{Concluding remarks}
\label{sec:7}
In this short paper we provided empirical evidence gathered from high frequency futures
data corresponding to various liquid futures assets that the impact (as measured from the return
variance or using the standard definition) of a trading order on the midpoint price depends
on intertrade duration. We have also shown that this property can also be observed at
coarser scale. A similar study of the spread value confirmed the idea that order books are less
filled when trading frequency is very high.
Notice that we did not interpret in any causal manner our findings, i.e., we do not assert that
a high transaction rates should be responsible for the fact that books are empty.
It just appears that both phenomena are highly correlated and this observation has to
be studied in more details. In a future work, we also plan to study the consequences of these observations
on models such those described in the introductory section (Eq. \eqref{Eq:TransactionTime}).
A better understanding of the aggregation properties (i.e., large values of $N$) and therefore of correlations
between successive trades will also be addressed in a forthcoming study.

\section*{Acknowledgments}
The authors would like to thank Robert Almgren (Quantitative Brokers and NYU) for many useful comments and support. They would also like to thank Mathieu Rosenbaum, Arthur Breitman and Haider Ali for inspiring discussions.

\bibliographystyle{plain} %style
\bibliography{the_nature_of_price2} %nom du fichier

\end{document}